\documentclass[aps,showpacs,twocolumn]{revtex4}
\newcommand{\dslash}{D\!\!\!\!\slash}
\usepackage{graphics}
\begin{document}

\iffalse
\begin{flushright}
DOE/ER/40762-329\\
UM PP\#05-021
\end{flushright}
\fi

\title {\boldmath{Large $N_c$ QCD at non-zero chemical potential}}

\author{Thomas D. Cohen}
\email{cohen@physics.umd.edu}

\affiliation{Department of Physics, University of Maryland,
College Park, MD 20742-4111}

\begin{abstract}
The general issue of large $N_c$ QCD at nonzero chemical
potential is considered with a focus on understanding the
difference between large $N_c$ QCD with an isospin chemical
potential and large $N_c$ QCD with a baryon chemical potential. A
simple diagrammatic analysis analogous to `t~Hooft's analysis at
$\mu=0$ implies that the free energy with a given baryon chemical
potential is equal to the free energy with an isospin chemical
potential of the same value plus $1/N_c$ corrections.
Phenomenologically, these two systems behave quite differently. A
scenario to explain this difference in light of the diagrammatic
analysis is explored.  This scenario is based on a phase
transition associated with pion condensation when the isospin
chemical potential exceeds $m_\pi/2$; associated with this
transition there is breakdown of the $1/N_c$ expansion---in the
pion condensed phase there is a distinct $1/N_c$ expansion
including a larger set of diagrams. While this scenario is
natural, there are a number of theoretical issues which at least
superficially challenge it. Most of these can be accommodated.
However, the behavior of quenched QCD which raises a number of
apparently analogous issues cannot be easily understood
completely in terms of an analogous scenario.  Thus, the overall
issue remains open.
\end{abstract}

%\pacs{11.15.Pg,11.10.Wx,11.30Rd}

\maketitle

\section{Motivation}

The problem of studying QCD at nonzero chemical potential is at
the crux of nuclear physics.  If  reliable methods could be
developed for doing this, then one could study cold nuclear matter
at zero pressure (which is not a bad caricature of the interiors
of real nuclei) as well as matter under extreme conditions which
may be relevant in astrophysics or heavy ion reactions.  The
ultimate purpose of this paper is to begin to explore the
implications of the large $N_c$ limit of QCD to this problem.
Large $N_c$ QCD has proven to be a useful tool to gain insights
into hadronic physics and thus it is reasonable to suppose that
it may prove similarly useful to study QCD with a chemical
potential\cite{Col}.

Before proceeding, it should be noted that the applicability of
large $N_c$ QCD to problems in the nuclear physics domain such as
nuclear matter may be quite problematic due to the relevant scales
in the problem.  In a formal large $N_c$ expansion one organizes
things only by factors $1/N_c$.  This is clearly useful if $N_c$
is large enough and if all of the coefficients multiplying the
$1/N_c$ factors are ``natural''.  However, if there are a wide
variety of scales in the problem which have an origin different
from $1/N_c$ physics, it may well be that the $1/N_c$ expansion
does not do a good job of organizing the physics into a hierarchy
of scales amenable to expansion.  In fact, in nuclear physics
many important scales are much smaller than typical hadronic
scales for reasons which have nothing to do with large $N_c$
physics.  For example, the binding of two nucleons is formally of
order $N_c^1$.  However,  the binding of the deepest bound
two-nucleon state is the deuteron at $\sim 2$ MeV. This is
two-order magnitude smaller than the nucleon-Delta mass splitting
($\sim 300$ MeV) which is of order $1/N_c$.  Thus one might
reasonably worry that the vastly different scales in the problem
presumably make the large $N_c$ expansion useless for
quantitative or semi-quantitative studies of nuclear physics.

Given this caution, one might conclude that it is pointless to
use large $N_c$ methods in the nuclear domain.  However, this
would be hasty.  In the first place, there may be problems for
which the scales present in nuclear physics do not mix strongly
with hadronic scales and which may lead to a useful expansion
quantitatively.  Studies of the nucleon-nucleon potential in
large $N_c$ show that the patterns of large and small
spin-isospin terms in the potential are consistent with results
from straightforward $1/N_c$ counting\cite{KMKS}.  Moreover, even
if the large $N_c$ expansion turns out to be of little utility
quantitatively it can be used to answer questions of principle.
For example, the question of whether the nucleon-nucleon force is
consistent with a meson-exchange picture can be studied in the
large $N_c$ context\cite{NN}.  The issue is nontrivial since
diagrams associated with multi-meson exchange are powers of $N_c$
larger than the allowed total value. However, cancellations occur
between classes of such diagrams to yield results consistent with
the standard $N_c$ counting. In a similar way one might hope to
gain qualitative insight into QCD at non-zero chemical potential.

This problem is of practical importance in nuclear physics. There
is also an underlying theoretical mystery. At zero temperature the
baryon density is strictly zero if the absolute value of the
chemical potential is less than the critical value to make nuclear
matter.  However, from the perspective of the functional integral
it is quite unclear {\it why} the density is zero.  This problem
has been dubbed the Silver Blaze problem\cite{SB,Kog} after the
Arthur Conan Doyle  story ``Silver Blaze'' in which the key clue
is a dog {\it not} barking in the night.  Resolving this problem
is of potential importance since understanding how nuclear matter
forms for chemical potential greater than this critical value
presumably depends on understanding how it doesn't form below. It
is plausible that large $N_c$ QCD can give insight into this
issue of principle.

One possible way to gain such insight into the problem of QCD at
a finite baryon chemical potential is to compare QCD with an equal
chemical potential for up and down quarks (which will yield
symmetric nuclear matter) to QCD with chemical potentials for up
and down quarks which are equal in magnitude but opposite in sign
(which makes ``isospin matter'', {\it i.e.}, a pion condensate at
low densities). One reason to focus on this comparison is that the
isospin matter case  is well understood  at low densities.
Firstly, at a phenomenological level this problem is amenable to
treatment using chiral perturbation theory \cite{SS}.  Secondly,
the analogous isospin Silver Blaze has been solved: the relevant
configurations which contribute to the functional integral have
functional determinants equal to their zero chemical potential
value for all chemical potentials less than the critical value
(which is a chemical potential per quark of $m_\pi/2$)\cite{SB}.
Thus if we can understand the difference between the baryon
chemical potential case and the analogous isospin case, we can
understand the baryon case.  At a formal level these two systems
are quite similar, but at a phenomenological level they are very
different and our goal is to understand this difference.

On first sight it seems plausible that large $N_c$ QCD might
provide import insights into the differences between the two
systems. Indeed, a simple diagrammatic analysis parallel to the
original analysis of `t Hooft (done for the zero chemical
potential case)\cite{Hoo} yields the result that the density of up
quarks depends only on the up quark chemical potential while the
density of down quarks depends only on the down quark chemical
potential plus corrections which are suppressed at large $N_c$.
This implies that baryon matter with a certain chemical potential
per quark has the same magnitude of quark density for up and down
quarks as isospin matter of the same chemical potential: the only
difference being that in the isospin case one of the flavors has
a negative density. A simple way to understand this is that large
$N_c$ QCD is quenched: in quenched QCD quark properties are only
affected by their propagators and not via the functional
determinants so there is no way for the chemical potential for one
flavor of quark to affect the behavior of a quark of a different
flavor. However, this is at odds with known phenomenology. At
zero temperature, the critical value for the isospin chemical
potential goes to zero in the chiral limit while the critical
value of the baryon chemical potential presumably does not. Thus,
a simple large $N_c$ analysis of the problem does not resolve the
problem of interest, rather it yields a new puzzle: namely, why
does the simple `t Hooft type analysis fail?

The immediate goal of this paper is to attempt to resolve this
problem:  before the general behavior of QCD at finite baryon
chemical potential can be understood, this issue must be dealt
with. This paper explores a scenario which appears to provide a
natural resolution to the problem---the $1/N_c$ expansion breaks
down due to infrared effects associated with spontaneous breaking
of U(1) isospin symmetry (associated with the third component)
which thereby invalidates the `t Hooft type diagrammatic analysis.
Despite its simplicity this explanation raises some interesting
and subtle theoretical issues. Indeed, on its face this
resolution immediately poses new problems. One problem concerns
the nature of the break down of the $1/N_c$ expansion. A Taylor
expansion breaks down when it enters a regime in which it ceases
to converge.  This does not appear to be the nature of the
breakdown here---one can be in a regime in which both the isospin
matter case and the baryon matter case are well described by
Taylor series in $1/N_c$ and yet they do not agree with each
other.  The fact that they are well described by a converged
series suggests that the expansion has not broken down, yet the
fact that they give distinct answers suggests that it has.  What
is going on?  A second problem involves the behavior of the
system at small, but nonzero temperatures and small chemical
potentials. In this regime one can use chiral perturbation theory
to compute the densities and by explicit computation it can be
seen that the isospin chemical case differs radically from the
baryon chemical potential case despite the fact that the large
$N_c$ analysis says they should both be equal at leading order.
This is apparently in contradiction to the overall resolution of
the paradox:  in this regime there is no phase transition and
hence no apparent reason why the $1/N_c$ expansion breaks down.
As will be discussed in detail in this paper these apparent
problems can also  be resolved and reconciled with the simple
explanation given above.

There is an additional problem with the scenario and that relates
to the quenched approximation to QCD. Lattice simulations show
that quenched QCD incorrectly gives the critical point for the
baryon chemical potential qualitatively (at a chemical potential
per quark of $m_\pi/2$)\cite{K}.  Since the diagrammatic analysis
implies that the large $N_c$ limit of QCD {\it is} quenched, the
known behavior of quenched QCD suggests that whatever problem
exists at large $N_c$, it is associated with the baryon number
case and not the isospin case. This seems to be at odds with the
general resolutions since it is the isospin chemical potential
which leads to spontaneous symmetry breaking.  There is a
plausible way to evade this difficulty---that quenched QCD
undergoes a phase transition at a chemical potential per quark of
$m_\pi/2$ associated with breaking a symmetry which is not
present in full QCD. In this picture, quenched QCD with a baryon
chemical potential behaves analogously to QCD with an isospin
chemical potential. However, as will be discussed in detail this
explanation is problematic when viewed at a diagrammatic level.

The difficulties associated with a diagrammatic description of
quenched QCD raises doubts about the scenario presented here. It
is plausible, however, that the problem is associated with
peculiarities of quenched QCD (which, after all is not even a
physical theory). If this is the case, then the explanation
outlined here may well be correct.

This paper is organized as follows.  In the next section, the
underlying theoretical differences between the two systems (QCD
with isospin chemical potential and QCD with a baryon chemical
potential) is briefly reviewed as is the phenomenology of the two
cases.   In the following section, a simple large $N_c$ analysis
based on diagrammatics is given which yields the prediction that
baryon matter with a certain chemical potential per quark has the
same magnitude of quark density for up and down quarks as isospin
matter of the same chemical potential.  This is the principal
problem addressed here. Next comes a section discussing how a
phase transition associated with pion condensation appears to
resolve this problem and various theoretical issues associated
with this resolution.  The final section discusses the challenges
to the scenario posed by quenched QCD.

\section{background \label{bg}}

For simplicity the problem will be studied in the limit of
unbroken isospin so the up and down quark masses will be taken as
degenerate and electromagnetic effects ignored. The problem is
most easily expressed in terms of QCD with a separate chemical
potential for up and down quarks; that is, consider a system whose
Lagrangian is given by
\begin{equation}
{\cal L}={\cal L}_{\rm QCD} + \mu_u \overline{u} u + \mu_d
\overline{d} d  \; .
\end{equation}
With Lagrangian one can evaluate the partition function---at
least in principle---and from the partition function one can
compute the free energy density:
\begin{equation}
{\cal G}(\mu_u,\mu_d,T) = -\frac{\left(\log(Z(\mu_u,\mu_d,T)\right
)} {\beta V}
\end{equation}
where $V$ is the spatial volume of the system and $\beta=1/T$.
The densities of the two species of quark are given by
\begin{equation}
\rho_u =-\frac{\partial {\cal G}}{\partial \mu_u} \; \; , \; \;
\rho_d=-\frac{{\partial \cal G}}{\partial \mu_d} \; .
\end{equation}
Two important cases can be distinguished---a baryon matter ({\it
eg.} nuclear matter) for which $\mu_u=\mu_d \equiv \mu$ and
isospin matter ({\it eg.}, a pion condensate for $\mu_u=-\mu_d
\equiv \mu$):
\begin{eqnarray}
Z_B(\mu,T) &\equiv& Z(\mu,\mu,T) \; \; , \; \; Z_I(\mu,T) \equiv
Z(\mu,-\mu,T) \nonumber \\
 {\cal G}_B(\mu,T) &\equiv& {\cal G}(\mu,\mu,T) \;
\; , \; \; {\cal G}_I(\mu,T) \equiv {\cal G}(\mu,-\mu,T)
\label{dif} \end{eqnarray} The focus of this paper is on the
difference between $ {\cal G}_B(\mu,T)$ and  ${\cal G}_I(\mu,T)$
in the large $N_c$ limit of QCD.

As seen in ref.~\cite{TDC},  both $Z_B(\mu,T)$ and $Z_I(\mu,T)$
can be represented as Euclidean space functional integrals:
\begin{widetext}
\begin{eqnarray}
Z_B(\mu,T) \,&  = & \, \int  d [A] \,\prod_{i=\rm heavy} {\rm
det}(\dslash + m_i) \,   \left ( \, {\rm det}(\dslash
+ m - \mu \, \gamma_0 \, ) \, \right )^2 e^{-S_{YM}} \nonumber \\
Z_I(\mu,T) \,&  = & \, \int  d [A] \, \prod_{i=\rm heavy} {\rm
det}(\dslash + m_i) \,{\rm det}(\dslash + m - \mu \, \gamma_0 \,
) \,  \, {\rm det}(\dslash + m + \mu \, \gamma_0 \, ) \,
e^{-S_{YM}} \nonumber \\
&= & \int  d [A] \, \left | \, {\rm
det}(\dslash + m - \mu \, \gamma_0 \, ) \, \right |^2 e^{-S_{YM}}
\; , \label{Zff}\end{eqnarray}
\end{widetext}
where YM  indicates the Yang-Mills action, ``heavy'' indicates
all quarks heavier than up or down and $m$ is the mass of the
light quarks. In these expressions the standard boundary
conditions are imposed: periodic for the gluons and anti-periodic
for the fermions with an extension in the time direction of
$\beta=1/T$. Note in the expression for $Z_B$ the functional
determinant is squared (one for each flavor) while for the first
form for $Z_I$ the chemical potential for the two flavors is
opposite.  The second form for $Z_I$ exploits the identity ${\rm
det}(\dslash + m + \mu \, \gamma_0 \, )^*={\rm det}(\dslash + m -
\mu \, \gamma_0 \, )$ which can easily be proved \cite{TDC}.

Comparing the second form for $Z_I$ with the form for $Z_B$ in
Eq.~(\ref{Zff})and the fact that $  {\rm det}(\dslash + m_i)$ is
real and positive, one sees that the only difference between the
two is the phase of the functional determinant in the integrand.
Thus, the entire difference between the phenomenology of nuclear
matter from isospin matter (a pion condensate) stems from this
phase. As observed in ref.~\cite{TDC,Kog} the fact that $Re \left
( \, {\rm det}(\dslash + m - \mu \, \gamma_0 \, ) \, \right )^2
\le \left | \, {\rm det}(\dslash + m - \mu \, \gamma_0 \, ) \,
\right |^2$ allows one to deduce in an {\it a priori} way that
$Z_I(\mu,T) \le Z_B(\mu,T)$. The issue of relevance in this paper
is whether large $N_c$ QCD allows us to deduce anything else
about the relationship between $Z_I(\mu,T)$ and $Z_B(\mu,T)$.

Before turning to the large $N_c$ limit it is useful to review
what is known about the phenomenology of baryon matter and isospin
matter. Here the focus will be on the zero temperature case.
There are two reasons for this focus.  The first is simply because
the Silver Blaze problem only arises at zero temperature.  The
second is the analysis in this paper will ultimately be done in
the large $N_c$ limit where one expects the general phenomena of
large $N_c$ continuum reduction to apply
\cite{lncr1,lncr2,lncr3,lncr4,cc}. However, as shown in
ref.~\cite{TDC2} large $N_c$ continuum reduction means that at
large $N_c$ the expectation of an operator will be independent of
temperature for all temperatures below some critical value.  Thus
it is sufficient to study the $T=0$ case to understand the full
low temperature behavior at large $N_c$.

First consider the case of isospin matter.  At low chemical
potential one expects that the system remains in the vacuum state
for chemical potentials less than $m_\pi/2$, which is the critical
point since the pion is the state with the smallest mass per unit
isospin in QCD. At $\mu=m_\pi/2$ there will be a second-order
phase transition to a dilute pion condensate whose density then
grows with $\mu$.  The system can be analyzed in chiral
perturbation theory\cite{SS} at tree level and the up and down
quark densities are given by
\begin{widetext}
\begin{equation}
\rho_u(\mu) = -\rho_d(\mu)= {\rm sign}\left ( \mu \right ) \,
\theta \left (4 \mu^2-m_\pi^2 \right) \, 2 \mu f_\pi^2 \left ( 1 -
\frac{m_\pi^4}{16 \mu^4} \right ) \left (1 + {\cal O}\left(
\frac{\mu^2 , m_\pi^2}{\Lambda^2} \right ) \right )  \label{phenI}
\end{equation}
\end{widetext}
where $\Lambda$ is a typical hadronic scale (for example
$m_\rho$).  From Eq.~(\ref{phenI}) it is apparent that if one can
neglect the higher order terms in the chiral expansion, then there
is a second-order transition as a function of $\mu$ with a
critical $\mu$ of $m_\pi/2$; there is no discontinuity in the
density at the critical point.  On very general grounds the
higher-order terms are not expected to destroy this qualitative
picture: there will be a second-order transition at $\mu=m_\pi/2$.

The case of baryon matter is quite different.  The transition as
a function of $\mu$ is first order at $T=0$.  Below a critical
value of the chemical potential the density is zero. Immediately
above this value the density is given by the saturation density
of nuclear matter $\rho_0 \approx .16 \, {\rm nucleons}/{\rm
fm}^3$. The critical chemical potential is given by
\begin{equation}
\mu_0 \equiv (M_N - B/A)/N_c  \label{mu0}
\end{equation}
where $N_c$, the number of colors, is three for the physical
world, $M_N$ is the nucleon mass and $B/A$ is the binding energy
per nucleon of saturated nuclear matter which is estimated to be
approximately 16 MeV. These properties of baryon matter can be
easily inferred from the extrapolation of the properties of finite
nuclei\cite{nm}.

It is apparent that the behavior of the two systems is completely
different.  The order of the transition is different as is the
critical chemical potential.  Perhaps more striking than the
numerical differences in the critical chemical potentials are the
qualitative ones: in the isospin case the critical chemical
potential is of a chiral origin and goes to zero as the quark
masses go to zero; this is not true in the baryon case.  Clearly,
the single sign difference in Eq.~(\ref{dif}) leads to profound
phenomenological differences.  In the following sections the issue
of what large $N_c$ QCD tells us about the way a sign plays out
dynamically will be investigated.

\section{Diagrammatic Analysis \label{dia}}

\begin{table}\label{T}
\begin{tabular}{|c|l|}
\hline  {\bf Order} & \hspace{.5in} {\bf Class of Diagrams} \\
\hline \hline
$N_c^2$ & $\bullet$ planar graphs with no quark loops. \\
\hline $N_c^1 $ &$\bullet$  planar graphs with one quark loop
that \\
& forms a boundary of the graph. \\
\hline $N_c^0$ & $\bullet$ graphs with one nonplanar gluon \\
& and no quark loops.  \\
 & $\bullet$ planar graphs with two nonconcentric quark \\
 & loops that form inner
 boundary of
 the graph.  \\
 \hline
\end{tabular}
\vspace*{.15in} \caption{$N_c$ counting of leading classes of QCD
diagrams from `t Hooft counting rules. All diagrams not in these
classes are lower order in the expansion.  The classes are
specified by topology. Note that there is some freedom in how
graphs are drawn. A graph is in a class listed in the table if it
{\it can} be drawn in the topology indicated.} \vspace*{.25in}
\end{table}

The study of large $N_c$ QCD began with `t Hooft's seminal work
\cite{Hoo}. The key theoretical tool  introduced in this was a
classification of perturbative QCD diagrams into classes
according to their $N_c$ scaling.  The assumptions underlying this
classification are very simple: the quarks are taken to be in the
fundamental representation of SU($N_c$) while the gluons are in
the adjoint; the coupling constant is taken to scale as
$N_c^{-1/2}$ while the quark mass is taken to scale as $N_c^0$.
These rules ensure a smooth large $N_c$ limit for mesonic
properties.   The analysis is greatly simplified by exploiting
the fact that color content of a color adjoint gluon is
essentially that of a fundamental-antifundamental bilinear with a
color singlet contribution removed.  However, this color singlet
combination is a $1/N_c^2$ effect so that at leading order and
next-to-leading order as far as color counting is concerned the
gluon acts like a quark-antiquark pair.  This is beautifully
encoded via the introduction of `t Hooft's double line
representation for the gluons representing the color flow (the
fundamental and anti-fundamental indices have flow in opposite
directions).  The quarks are represented by single lines
representing the flow of the color in the fundamental
representation.  Color conservation requires that the color
entering a vertex must leave it; thus all color lines eventually
form closed loops.  The $N_c$ counting of a diagram is then given
by a factor of $N_c$ for each closed color loop and a factor of
$N_c^{-1/2}$ for every coupling constant.

The results of this analysis are well known and summarized in
Table \ref{T}.  The issue of what modifications are needed to this
analysis of the counting rules when a chemical potential is
included is critical here.  However, it can be dealt with
trivially: formally the analysis of the scaling goes through
without modification, provided that the chemical potential is of
order $N_c^0$.  The reason for this is straightforward: the
chemical potential enters a diagram only via the quark
propagator.   If the quark mass and the chemical potential are
each of order $N_c^0$, then the quark propagator itself is also
of order $N_c^0$.  Since the inclusion of a chemical potential of
order unity does not alter the $N_c$ dependence of the
propagator, its inclusion does not alter the $N_c$ dependence of
the diagrams and Table \ref{T} continues to apply.

In order to use this topological classification of diagrams to
deduce the $N_c$ scaling properties of physical observables, an
additional assumption must be made. Namely, that the leading
$N_c$ for some quantity is fixed by the $N_c$ counting of the
lowest-order class of graphs which can contribute to the quantity
in a perturbative expansion.  This assumption is nontrivial since
a theory need not be equivalent to the perturbative expansion
summed to all orders; intrinsically nonperturbative phenomena can
contribute and {\it a priori} there is no guarantee that
nonperturbative contributions will follow the same $N_c$ scaling
laws as the perturbative ones.  However, this assumption has
worked well for the case of zero chemical potential in the sense
that there are no known cases where intrinsically nonperturbative
contradictions spoil the $N_c$ scaling based on diagrammatics. It
seems reasonable as a tentative first step in the study of
nonzero chemical potentials to make the same assumption.

The free energy is simply the sum of all connected diagrams
(assuming that the resummed perturbative expansion is equivalent
to the theory). The dependence of the free energy on the chemical
potential requires the contribution of at least one quark loop.
From Table \ref{T}, it is apparent that the leading class of
diagrams containing at least one quark loop are the one-loop
diagrams where the quark loop bounds the diagram and contributes
at order $N_c^1$; two-quark-loop contributions are at least one
order down in $N_c$.  Since the gluons are flavor neutral,
one-quark-loop diagrams are either loops with up quarks or down
quarks (the contributions of heavier quark loops are neglected
since the focus is on the dependence of the up and down quark
chemical potential.)   This implies that the $\mu$-dependent part
of the free energy density at leading order in the expansion is
given by
\begin{equation}
{\cal G}(\mu_u,\mu_d,T)= N_c \left (f(\mu_u,T) + f(\mu_d,T)\right
) \left ( 1 + {\cal O}(1/N_c) \right )\;. \label{form}
\end{equation}
The first term on the right-hand side is the sum of the
one-up-quark loop contributions while the second term represents
the one-down-quark loops.  Note that functions $f(\mu )$ for the
up-quark and down-quark contributions are identical due to
isospin invariance.  The overall factor of $N_c$ follows from the
scaling rule seen in Table \ref{T}.  It is easy to see why
Eq.~(\ref{form}) must hold---the large $N_c$ limit of QCD is
quenched. This implies that the contribution of the two flavors
must be independent---the up quark density, for example, cannot
depend on the down quark chemical potential.  This independence
is encoded in Eq.~(\ref{form}).

The apparent phenomenological consequences of  Eq.~(\ref{form})
are quite stark. Combining eqs.~(\ref{form}) and (\ref{dif})
yields
\begin{eqnarray}
{\cal G}_I(\mu,T)& = & N_c \left (f(\mu,T) + f(-\mu,T)\right )
\left ( 1 + {\cal O}(1/N_c) \right ) \nonumber
\\
{\cal G}_B(\mu,T)& = &2N_c f(\mu_u,T) \left ( 1 + {\cal O}(1/N_c)
\right ) \; .\label{B}
\end{eqnarray}
It is easy to see that $f(\mu,T)$ must be an even function of
$\mu$ if the $1/N_c$ expansion holds. ${\cal G}_B$ is given by
Eq.~(\ref{B}). However,  CPT implies that ${\cal G}_B(-\mu,T)=
{\cal G}_B(\mu,T)$ since anti-nuclear matter has the same mass as
nuclear matter. This immediately implies that if $1/N_c$
corrections can be neglected, then $f(\mu,T)=f(-\mu,T)$. However,
if $f$ is even, then Eq.~(\ref{B}) implies that
\begin{equation}
{\cal G}_I(\mu,T) =  {\cal G}_B(\mu,T) \left ( 1 + {\cal
O}(1/N_c) \right ) \; .\label{para}
\end{equation}
Equation (\ref{para})  states that the free energy of baryonic
matter is identical to that of isospin matter at the same
chemical potential plus $1/N_c$ corrections.  It is at  the core
of the paradox: as discussed in Sec.~\ref{bg}, at a
phenomenological level the two systems are radically different
yet the $1/N_c$ expansion suggests that their differences should
be small.

\section{A possible resolution \label{PR}}

The resolution to this paradox, presumably either resides in the
theoretical analysis or in the phenomenological description.

Superficially, the logic underlying the analysis in the previous
section was identical to that used by `t Hooft in the zero
chemical potential case where the analysis does appear to be
reliable. Thus, it is prudent to first focus on the possibility
that phenomenology as being the culprit. {\it A priori} this may
not seem to be  too implausible. After all, as discussed in the
introduction, the large $N_c$ limit of QCD can easily fail for
nuclear phenomena given the very small characteristic scales
involved. Thus, one might hope that Eq.~(\ref{para}) holds
formally;  this would imply that the fundamental problem is
simply that the $1/N_c$ corrections swamp the leading term. This
would make nuclear matter very different from isospin matter. This
resolution to the paradox amounts to the assertion that if we
only lived in a truly large $N_c$ world (say $N_c=1001$ rather
than $N_c=3$),  isospin matter and nuclear matter at the same
chemical potential would have virtually the same density.

However, this phenomenological  resolution to the paradox is
almost certainly not correct.  Note that for the isospin case
there is a second-order transition at $\mu=m_\pi/2$ for $T=0$. For
Eq.~(\ref{para}) to hold in a large $N_c$ world the critical
chemical potential for the baryon case would also have to be
$\mu=m_\pi/2$ and thus tend to zero in the chiral limit.  For
this to occur one would have to make nuclear matter which is so
deeply bound that the binding energy between nucleons cancels the
nucleon's mass. This is quite far fetched since the nucleon mass
appears to have nothing to do with the chiral limit.  It is even
more far fetched when one considers the second-order nature of
the transition. The density at $\mu$ just above the transition in
the isospin case can be made arbitrarily small so that
Eq.~(\ref{para}) implies that the baryon density would also be
arbitrarily small just above the transition. However, if the
density in the baryon case is dilute then the nucleon-nucleon
interactions would be small and thus could not cancel the nucleon
mass.

Having disposed of the phenomenological resolution, one is forced
to conclude that Eq.~(\ref{para}) is wrong and, hence, so is the
theoretical justification underlying it.  The issue then becomes
identifying what is wrong with the standard $1/N_c$ analysis of
the diagrams, which as we have noted is identical to that used by
`t Hooft at zero chemical potential. In attempting to resolve
this theoretically there are a number of issues that appear to
complicate the analysis.

The first is that there is no obvious breakdown of the $1/N_c$
expansion.  If we concentrate on ${\cal G}_B(\mu,T)$ and ${\cal
G}_I(\mu,T)$ and imagine writing them both as Taylor series in
$1/N_c$, there is no indication that the series  breaks down for
either quantity.  Consider, for example, the regime $T=0$, $\mu_0
\gg \mu > m_\pi/2$, where $\mu_0$ is given in Eq.~(\ref{mu0}). In
this regime the  ${\cal G}_B(\mu,T)$ is strictly zero. Thus
${\cal G}_B(\mu,T)$ is accurately described by a Taylor series in
$1/N_c$ with all terms zero.  On the other hand, ${\cal
G}_I(\mu,T)$ is finite and for small enough values of $\mu$ and
$m_\pi$ can be arbitrarily well described by Eq.~(\ref{phenI}).
Since this regime is supposed to be the regime of validity of
chiral perturbation theory, one expects that a systematic series
in pion loops is possible since the chiral expansion organizes
into a loop expansion for pion contributions.  However, the loop
expansion for meson is also the $1/N_c$ expansion and hence one
expects that if the chiral expansion is legitimate so is the
$1/N_c$ expansion.  This suggests that, at least in this regime,
a $1/N_c$ expansion is valid for both quantities of interest.
This poses a problem since we  expect that the most plausible
theoretical explanation of the difference between ${\cal G}_I$
and ${\cal G}_B$ is due to a breakdown  of the $1/N_c$ expansion;
if the expansion does not break down, they must agree up to
$1/N_c$ corrections.

A second issue is associated with the  regime of low temperatures
and small chemical potentials: $\mu, T \ll m_\pi$ where, at least
formally $\mu$, $T$ and $m_\pi$ are taken to be of order
$N_c^0$.  In this regime, chiral symmetry implies that the system
behaves as a gas of  weakly interacting pions from which one can
directly compute the free energy density of isospin matter:
\begin{eqnarray}
{\cal G}_I(\mu,T) & = &f(T) \left (1 + 2 \cosh \left ( \mu/T
\right )
\right ) + {\cal O} \left ( e^{- 2 m_\pi/T} \right ) \nonumber \\
f(T) & = & T \int \frac{{\rm d}^3 k}{{2 \pi}^3} \log \left ( 1 -
e^{-\sqrt{k^2 + m_\pi^2}/T }\right ) \; \label{r2}\end{eqnarray}
Thus we see that the free energy density in this regime, though
small, is formally of order $N_c^0$.  This should be contrasted to
the baryon free energy which, in this regime, is exponentially
small in a $1/N_c$ sense ${\cal G}_B(\mu,T) \sim e^{-Nc}$. Thus,
the two free energies continue to differ in this regime.  The
significance of the behavior in this regime becomes clear if one
studies ${\cal G}_I(\mu,T)$ as a function of $\mu$ at fixed but
small $T$. Equation (\ref{r2}) shows explicitly that in this
regime there is no discontinuity as a function of $\mu$ all the
way down to $\mu=0$; in the regime under consideration there is
no phase transition to a pion condensate. However, since the
problem of isospin matter not matching baryon matter persists in
this regime it suggests that the phase transition is not the
origin of the problem.

These issues appear to suggest that the problem is not associated
with a breakdown  of the $1/N_c$ expansion for the free energies,
and is not associated with a discontinuity of the free energy as a
function of $\mu$.  However, it will be argued here that the cause
of the problem is with the description of the isospin matter
system, and is caused by a breakdown of the $1/N_c$ expansion
which, in turn, is associated with a  phase transition.

Recall the crux of the problem.  A diagrammatic analysis at large
$N_c$ implies that at leading orders the two quark flavors do not
``talk to'' each other: down quark observables do not depend on
the up quarks and vice versa. If they do depend on each other at
leading order, than the expansion has broken down. This is
precisely what occurs for the case of isospin matter.  To see
this, let us  return to the regime with $T=0$ and $\mu > m_\pi/2$
and  $\mu, m_\pi \ll \Lambda$ is a typical hadronic scale.   In
this regime one can directly compute properties of isospin matter
using a lowest-order Lagrangian working at tree level.  This
analysis was originally carried out in ref.~\cite{SS} and was the
basis for Eq.~(\ref{phenI}).  The important thing here is that in
this regime there is a pion condensate which can be computed
easily. Expressed in terms of the up and down quark fields it is
given by:
\begin{widetext}
\begin{eqnarray}
\langle i m \overline{u} \gamma_5 d \rangle & = & e^{i \phi}\,
\theta \left ( |\mu| - m_\pi/2  \right ) \, \frac{m_\pi^2
f_\pi^2}{2}
 \, \sqrt{1 -\frac{m_\pi^4}{16 \mu^4} } \, \left ( 1 + {\cal O}
\left( \frac{m_\pi^2, \mu^2}{\Lambda} \right ) \right ) \; ,
\nonumber
\\ \langle i m \overline{d} \gamma_5  u  \rangle & = &
e^{-i \phi} \, \theta \left ( |\mu| - m_\pi/2  \right )\,
\frac{m_\pi^2 f_\pi^2}{2} \,  \sqrt{1 -\frac{m_\pi^4}{16 \mu^4}
}\,  \left ( 1 + {\cal O} \left( \frac{m_\pi^2, \mu^2}{\Lambda}
\right ) \right ) \; .\label{cond}
\end{eqnarray}
\end{widetext}
The phase in Eq.~(\ref{cond}) is arbitrary.  The presence of an
arbitrary phase is a signature of spontaneous symmetry
breaking---in this case the breaking of a U(1) symmetry
associated with the third component of isospin.  Of course, the
phase is uniquely fixed if by adding an infinitesimal symmetry
breaking term to the Lagrangian of the system.  This symmetry
breaking is due to an instability in the infrared.  This can be
seen by looking at the chiral susceptibility,
\begin{equation}
\chi_\chi \equiv \int {\rm d}^4 {\bf x} \,\left  \langle (i m
\overline{q} \gamma_5 \tau_x q)({\bf x})\, (i m \overline{q}
\gamma_5 \tau_x q)({\bf 0}) \right \rangle \label{chis}
\end{equation}
where the expression is for Euclidean space. In the broken phase
$\mu > m_\pi$ , $\chi_\chi$  is divergent while it is finite for
$\mu<m_\pi$. Regardless of the particular value of the phase,
Eq.~(\ref{cond}) implies that there is a nonvanishing expectation
value of an operator which connects up and down quarks.  This is
an unambiguous indication that the up quark and down quarks
``talk to'' each other in isospin matter, contrary to the large
$N_c$ analysis based on diagrams.  This shows that the problem
does indeed lie with the isospin matter and not with baryon
matter and is associated with the phase transition due to
spontaneous symmetry breaking.

\begin{figure}\label{loops}

\includegraphics{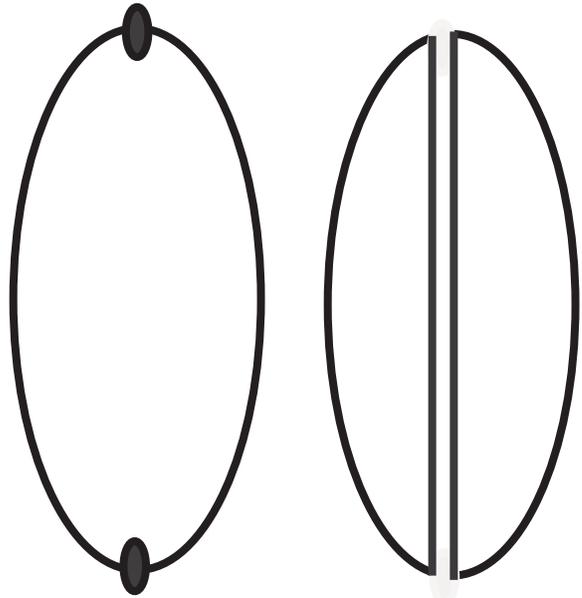}
\caption{Feynman graphs with (planar) gluon lines suppressed.
Left graph is a one-quark-loop graph with two insertions of the
operator $i m \overline{q} \gamma_5 \tau_x q$ (represented by
filled ellipses). This graph corresponds to the chiral
susceptibility $\chi_\chi$ defined in Eq.~(\ref{chis}).  Right
graph removes these insertions and replaces them with a
quark-antiquark pair carrying the same quantum numbers.}
\vspace*{.15in}
\end{figure}

It is easy to see at a diagrammatic level why the symmetry
breaking leads to a breakdown  of the naive $1/N_c$ expansion.
Consider the $\chi_\chi$ from Eq.~(\ref{chis}) which can be
understood as a quark loop with two insertions of the operator $i
m \overline{q} \gamma_5 \tau_x q$.  The fact that $\chi_\chi$
diverges in the broken phase indicates that even an
infinitesimally weak perturbation with these quantum numbers has
a finite effect and that this finite response occurs at leading
order in the $1/N_c$ expansion. Now suppose one were to replace
the two insertions by a quark-antiquark  pair connecting the
points as in Fig.~\ref{loops}.  Note that the graph with the
additional quark-antiquark pair creates an additional quark loop
and hence is formally down compared to a single quark loop by a
factor of $N_c$.  However, if the quark-antiquark pair carries
the quantum numbers of a pseudoscalar with the x component of
isospin, it acts on the original quark loop precisely in the same
way as two $i m \overline{q} \gamma_5 \tau_x q$ insertions and,
hence, in the broken phase can make a leading order contribution.

This explanation is both simple and natural.  However, it appears
to be in conflict with the arguments presented above.  Thus, it is
important to understand just how these arguments can be evaded.

The first issue was the lack of an obvious breakdown  of the
$1/N_c$ expansion for the quantities of interest---namely, the
free energy densities.  In the regime, $\mu_0 > \mu > m_\pi/2$,
both ${\cal G}_I$ and ${\cal G}_B$ have  well-defined Taylor
expansions in $1/N_c$.  How then can one argue that a breakdown
in the $1/N_c$ expansion for ${\cal G}_I$  to cause it to differ
from ${\cal G}_B$ as implied by the diagrammatic $1/N_c$
expansion?

The answer lies in the possibility that there are multiple
regimes in the problem, each with a  valid $1/N_c$ expansion but
which are not connected with each other due to a breakdown  of the
$1/N_c$ expansion at the boundary of the regimes. Specifically,
the scenario  is that there is a well-defined Taylor expansion in
$1/N_c$ for ${\cal G}_I$ in the regime $\mu <m_\pi/2$ and another
well-defined but disconnected expansion for $\mu > m_\pi/2$. In
this scenario, the regime for $\mu < m_\pi/2$ behaves according
to the large $N_c$ behavior as deduced from the diagrammatic
analysis ({\it i.e.}, ${\cal G}_I={\cal G}_B$ plus possible higher
order corrections). In the regime where $\mu > m_\pi/2$ (for
$T=0$), ${\cal G}_I$ still possesses a $1/N_c$ expansion but it is
a different  expansion than that given by the  diagrammatic
analysis of `t Hooft.  This is possible only if there is no smooth
way to connect the series for the two regimes due to a breakdown
of the expansion on the boundary $\mu=m_\pi/2$.

As it happens, the series in $1/N_c$ for ${\cal G}_I$  {\it does}
break down  at the boundary between these two regimes.  Note from
Eq.~(\ref{phenI}), that the expression contains a smooth function
of $\mu$ times a step function at $\mu^2=m_\pi^2/4$. Recall,
however, that the pion mass itself is a function of $N_c$:
\begin{equation}
m_\pi^2 = {m_{\pi}^2}_{(0)} + \frac{{m_\pi^2}_{(1)}}{N_c} +
\frac{{m_\pi^2}_{(2)}}{N_c^2} + .... \; , \label{piexp}
\end{equation}
where the coefficients ${m_\pi^2}_{(n)}$ is the $n$th term in the
expansion. The step function can be formally expanded in $N_c$
via insertion of the series in Eq.~(\ref{piexp}) and a subsequent
Taylor expansion.
\begin{widetext}
\begin{equation}
\theta \left ( 4 \mu^2 - m_\pi^2 \right )=  \theta \left ( 4 \mu^2
- {m_\pi^2}_{(0)} \right )\, - \,
  \frac{{m_\pi^2}_{(1)} \,
\delta \left ( 4 \mu^2 - {m_\pi^2}_{(0)} \right ) }{N_c } + {\cal
O}(1/N_c^2)
\end{equation}
\end{widetext}
The series clearly breaks down at the point
$\mu^2={m_\pi^2}_{(0)}/4 $ since the second term is a divergent
$\delta$ function.  Thus, the scenario in which there is a
well-defined $1/N_c$ expansion in the symmetry broken phase---but
a different $1/N_c$ expansion than in the unbroken phase---is
viable.

Next, let us consider the second issue, namely, the behavior of
the system in the  regime of low temperatures and small chemical
potentials, {\it i.e.} $\mu, T \ll m_\pi$.  The apparent problem
arises since formally $\mu$, $T$ and $m_\pi$ are of order
$N_c^0$.  Equation (\ref{r2}), however, shows that in this regime,
isospin matter has a finite free energy density as $N_c
\rightarrow \infty$ while baryon matter has zero free energy
density. This is in apparent conflict with the large $N_c$
expectation based on diagrammatics seen in Eq.~(\ref{form}).
However, it is also clear that the problem is not associated with
a phase transition to a pion condensed phase since no such
transition occurs in this regime. This, in turn, appears to
contradict the claim made above that the failure of the large
$N_c$ analysis is due to such a phase transition.

In fact, there is no contradiction.   While it is true that
isospin matter and baryon matter do differ in this regime, the
differences---although of relative order 100\%---are, in fact,
subleading in the $1/N_c$ expansion.  The standard $1/N_c$
counting yields a number density and a free energy density of
order $N_c^1$ for chemical potentials of order unity. This can be
seen straightforwardly from the `t Hooft analysis---the leading
order is one loop and hence proportional to $N_c$.  This behavior
is seen explicitly, for example, in Eq.~(\ref{phenI}): the
standard scaling rules $f_\pi^2 \sim N_c$, $ m_\pi^2 \sim N_c^0$,
$\mu \sim N_c^0$ yield densities and free energy densities of
order $N_c^1$ when they are non-zero. However, the same scaling
rules applied to Eq.~(\ref{r2}) gives ${\cal G}_I \sim N_c^0$.
Thus, in the regime considered here, both the isospin case and the
baryon case have free energies which are zero at order $N_c^1$.
They differ from each other at order $N_c^0$ ({\it i.e.}, at the
first subleading order) which is exactly what one expects from the
diagrammatic analysis.

The behavior seen here in going from zero to  finite temperature
with no effect on the leading order large $N_c$ behavior
illustrates a rather general feature of large $N_c$ QCD and is
related to the general phenomena of large $N_c$ continuum
reduction\cite{lncr1,lncr2,lncr3,lncr4,cc,TDC2}.  The  unusual
feature in the case of the free energy density of isospin matter
is that at zero temperature the observable is strictly zero
while  at non-zero temperature the observable is non-zero but
still subleading.

While the explanation given in this section here is rather natural
there remain some important open questions associated with this
scenario. Since the essence of the original paradox was seen in a
diagrammatic analysis, a complete resolution of the issue requires
an understanding of the issue in terms of diagrams. However, the
resolution to the problem as given above---while very
plausible---does not make clear precisely which class of diagrams
play a role at leading order in $1/N_c$.

Consider, for example, QCD with an isospin chemical potential.
The scenario outlined here---that there are two distinct $1/N_c$
expansions, one valid for $\mu<m_\pi/2$ and one for $\mu>m_\pi/2$
with the expansions breaking down for $\mu=m_\pi/2$---should be
describable in terms of diagrams.  The diagrams which contribute
at large $N_c$ in the regime $\mu<m_\pi/2$ are presumably the
same set which occur for $\mu=0$: one quark loop graphs with
planar gluons and the quark loop bounding the graph.  However,
above the phase transition our explanation requires a larger set
of graphs to contribute at leading order.  The argument associated
with Fig.~\ref{loops} implies that in the broken phase a divergent
$\chi_\chi$ implies that even infinitesimal perturbations with
the appropriate quantum numbers can lead to leading-order
response.  Additional quark loops can couple with these quantum
numbers; although the overall strength associated with adding a
loop is suppressed by $1/N_c$ they can nevertheless yield leading
order results due to infrared enhancements. There is, however, an
important open question associated with this, namely, just which
class of diagrams does contribute at leading order in this
regime?  A plausible conjecture is that all graphs with planar
glue and with an arbitrary number of closed quark loops bounding
forming an inner boundary of the graph ({\it i.e}, with no glue
inside the graph).  This is plausible in that it builds in all of
the standard $1/N_c$ rules that hold generally, except for the
quark loop rule (which we know to be violated in this regime).

A parenthetical remark may be useful here: the leading order
graphs containing quarks for $\mu=0$ are conventionally expressed
as single quark loops with all of the (planar) glue on the
inside. However, as Feynman graphs these are identical to single
quark loops with all of the (planar) glue on the outside.  The
important point is simply that they bound the graph.  However, if
one wants to generalize to a situation with multiple quark loops,
the only topological way to have all of the loops bound the graph
is to have the glue on the exterior of the quark loops.

It is not immediately apparent, however, if this conjecture is
correct.  Nor is there an obvious and straightforward way to
verify it.  Thus, determining whether the conjectured class of
diagrams for this regime is  correct is an important open
question.

A second open question concerns the regime of finite baryon
chemical potential.  The explanation given here suggests that the
standard diagrammatic treatment for large $N_c$ QCD as done for
$\mu=0$ (planar graphs with a single quark loop bounding the
graph) remains valid at least up to the phase transition where
nuclear matter is formed (which occurs at $\mu=\mu_0$).  What
class of diagrams contributes in the nuclear matter regime?  At
present we do not know.

Indeed, the general question of how QCD nuclear matter behaves at
large $N_c$ is at present unanswered.  We do not know, for
example, if the transition as a function of $\mu$ at $T=0$ is
first order (as in the case of $N_c=3$) or second order.  The
need to develop reliable theoretical tools to study QCD with a
baryon chemical potential near the transition to nuclear matter
is paramount to any attempt in understanding nuclear physics from
QCD. One may hope that large $N_c$ QCD will ultimately provide
insight into this problem, but at present we lack the tools even
to study large $N_c$ QCD in this regime.

To summarize this section:  a natural explanation of the different
behavior of large $N_c$ QCD with a baryon chemical potential from
large $N_c$ QCD with an isospin chemical potential is due to a
breakdown  of the $1/N_c$ expansion for QCD with an isospin
chemical potential at the point where a pion condensation phase
transition occurs.  The two issues raised in this section which
suggested difficulties for such an explanation are apparently
reconciled with it.  However, there remain open questions of how
the system ought to be described diagrammatically in the symmetry
broken regimes.

\section{Implications from Quenched QCD}

The behavior of quenched QCD with  a baryon chemical potential
raises issues in many ways quite analogous to those of QCD with
an isospin chemical potential.  Thus, if the scenario given in the
previous section is correct, one would expect a similar scenario
to apply for quenched QCD.  As we will see in this section, the
behavior of quenched QCD raises challenges to this scenario.

Recall the essence of the argument of Sec.~\ref{dia} that QCD is
quenched at leading order large $N_c$, at least in a diagrammatic
analysis.  However, it is long been known that quenched QCD at
non-zero baryon chemical potential gives perverse results.  In
lattice studies of quenched QCD,  the critical baryon chemical
potential was found to be at $m_\pi/2$ and not at
$\mu_0$.\cite{Kog} As $\mu$ exceeds $m_\pi/2$ there is evidence
that the chiral condensate changes nonanalytically and that the
baryon density discontinuously increases from zero. Thus a
description of baryon matter based on quenched QCD with a baryon
chemical potential appears to be completely wrong: in full QCD
the chiral condensate remains at its vacuum value and the baryon
density remains zero until $\mu=\mu_0$. This immediately poses a
challenge to the explanation in Sec.~\ref{PR} since it suggests
that the problem lies with the treatment of baryon matter and not
with isospin matter.

To see why, note that the qualitative difference between quenched
QCD and full QCD at non-zero chemical potential may be viewed as
another symptom of the same basic problem seen in the comparison
of baryon matter and isospin---a diagrammatic large $N_c$ analysis
fails. One can consider both full and quenched QCD at any $N_c$.
Presumably the two descriptions remain qualitatively different at
all $N_c$ with the transition for quenched QCD occurring at
$\mu=m_\pi/2$ and full QCD well above this (since there is
nothing special about $N_c=3$ where lattice simulations for
quenched QCD were done). On the other hand, a diagrammatic
analysis says that the baryon density in large $N_c$ QCD should
become equivalent to the density in quenched large $N_c$ QCD
since all extra quark loops cost a suppression factor of
$1/N_c$.  The fact that the two differ, seems to be completely
analogous to the fact that baryon matter and isospin matter
differ despite being equivalent at the diagrammatic limit.
Clearly, isospin matter plays no role in this problem and thus
the origin of the difficulty here cannot be a phase transition in
isospin matter.  Since the original problem (an isospin chemical
potential versus a baryon chemical potential) and this one (full
versus quenched QCD with a baryon chemical potential) have one
thing in common---full QCD with a baryon chemical potential---it
is natural to guess that the problem lies there.

The behavior of quenched QCD might be reconciled with our
explanation of the difference between isospin matter and baryon
matter if quenched QCD undergoes a phase transition at finite
baryon chemical potential in a manner analogous to the breaking in
isospin matter. However, this transition cannot be associated
with the breaking of a  standard baryon number U(1) symmetry,
since then one would be faced with the questions of why the
transition occurs at $m_\pi/2$ and why the transition does not
occur in full QCD.

If one views quenched QCD as merely a prescription to evaluate
functional integrals with the functional determinant set to
unity, it is hard to see how this possibility could be realized.
However, this prescriptive interpretation of quenched QCD has a
major drawback: the physical quantities of interest are not
obtained by differentiating a generating functional with respect
to sources---as one does with legitimate quantum field theories.
If one formulates  quenched  QCD in terms of a field theory with
a generating functional then one can see how such a symmetry
breaking could come about. To do this one starts with full QCD
and then adds a new set of particles: for each quark one adds an
associated bosonic quark (a colored spin-1/2 particle with bosonic
commutation relations) with the same mass and chemical potential
as the original quarks\cite{Mor}. These boson quarks, are
generally denoted as $\tilde{q}$.  Of course, since these fields
violate the spin and statistic theorem, their inclusion leads to
some clearly unphysical consequences such as states with negative
norm. Although the theory is physically nonsensical in these
ways, one can treat it as a problem in mathematical physics still
computed in the usual way. The bosonic quarks can be formally
integrated out of the functional integral and contribute a
factor  ${\rm det}(\dslash + m - \mu \, \gamma_0 \, )^{-1}$  to
the integrand of the gluonic functional integral while the
corresponding quark contributes a factor ${\rm det}(\dslash + m -
\mu \, \gamma_0 \, )$.  The product cancels and thus the usual
quark fermion determinant is replaced by unity---this is standard
quenching.

The key issue here is that the inclusion of bosonic quarks
implies  new symmetries for the system and such symmetries can
break leading to a phase transition. If quenched QCD  does undergo
a phase transition (associated with the breaking of one of these
new symmetries) when the baryon chemical potential reaches
$m_\pi/2$,  the difference between full and quenched QCD at large
$N_c$ appears to be easy to understand. In essence, quenched QCD
with a baryon chemical potential then behaves analogously to full
QCD with an isospin chemical potential: the naive $1/N_c$
expansion breaking down at the transition point. It was suggested
by Stephanov on the basis of a random matrix model study that one
expects a phase transition associated with the breaking of a
symmetry not present in full QCD\cite{Step}. Moreover, from first
principles one expects quenched QCD to undergo a phase transition
at $\mu=m_\pi$.  The theory has more degrees of freedom than the
usual QCD---it has boson quarks as well as ordinary quarks.
Consider an excitation created by the current
\begin{equation}
J=\overline{u} i \gamma_5 \tilde{u} + \overline{d} i \gamma_5
\tilde{u} \; . \label{J}
\end{equation}
The lightest state in the spectrum with these quantum numbers {\it
i.e.} the lightest state obtained by $J$ acting on the vacuum
might be called a ``baryo-pion''---it carries nonzero baryon
number but in many respects acts like a pion.   In particular,
one expects a phase transition at the same $\mu$ (for a baryon
chemical potential) as ordinary pions will condense (for an
isospin chemical potential).

To see this,  consider the contribution to the susceptibility for
the current $J$, namely,
\begin{equation}
\chi_J \equiv \int {\rm d}^4 x \, \langle J^\dagger(x)
J(0)\rangle\, ,
\label{chiJ}
\end{equation}
from  a one fermion loop diagram in quenched QCD with baryon
chemical potential $\mu$ and with some particular arrangement of
gluon lines.  It is easy to see from the Feynman rules that the
contribution is identical to the contribution to $\chi_\chi$, the
chiral susceptibility defined in Eq.~(\ref{chis})  at one fermion
loop (with an isospin chemical potential $\mu$) and with the same
arrangement of gluon lines (modulo an overall multiplicative
constant). The standard $1/N_c$ expansion holds for isospin
matter in the regime $\mu < m_\pi/2$, so the one fermion loop
planar graphs should correctly describe $\chi_\chi$ for large
$N_c$ and $\mu>m_\pi/2$. This means that in this regime $\chi_J$
is proportional to $\chi_\chi$.  Since $\chi_\chi$ diverges as
$\mu \rightarrow m_\pi/2$, so does $\chi_J$.   This type of
divergence typically signals the onset of a phase transition.
Physically one might think of such a transition as the
condensation of ``baryo-pions''.

Thus, quenched QCD does have an instability  at a (baryon)
chemical potential of $m_\pi/2$ of the sort typically associated
with a phase transition. This instability is presumably
associated with the breaking of a symmetry which is not present
in full QCD. The appearance of such a phase transition is what is
needed to reconcile the difference of behavior of quenched and
full  QCD with baryon chemical potential with our overall
explanation.

While this state of affairs is encouraging, there remains a
troubling unresolved issue: It provides no way to understand the
difference between full and quenched QCD with a baryon chemical
potential at large $N_c$ from a diagrammatic perspective.

Recall that our explanation for the different behavior was that
quenched QCD undergoes a phase transition at $\mu=m_\pi/2$
associated with the breaking of a U(1) symmetry which is not
present in the full theory.  Regardless of the phenomenology, we
know that the baryon density in quenched QCD is generically given
diagrammatically as the sum of one quark loop graphs with a single
insertion of the operator $\overline{q} \gamma_0 q$ on that loop.
Indeed, the essence of quenching is precisely that all quark
loops not connected to external currents do not contribute; they
are exactly canceled by bosonic quark loops. In the large $N_c$
limit, only  the planar graphs with the quark line bounding the
graph contribute.

A potential problem occurs when we consider the regime $\mu <
\mu_0$, where full QCD has not undergone a phase transition.
Since full QCD is in the same phase as at nonzero $\mu < \mu_0$ as
at $\mu=0$, one expects that the naive $1/N_c$ analysis based on
diagrams to hold in this regime: at leading order, the baryon
density is given by planar gluons with a single quark loop which
bounds the graph; an insertion of $\overline{q} \gamma_0 q$ is
attached to the quark line. Indeed, the scenario outlined in
Sec.~\ref{PR} is based on this assumption. However, this is
precisely the same set of diagrams that occurs in a calculation
of the baryon density in quenched QCD at large $N_c$ regardless
of the value of $\mu$. Thus, in the regime $m_\pi/2 < \mu <
\mu_0$, full and quenched QCD at large $N_c$ appear to have
exactly the same diagrams contributing, yet to predict different
densities. The difficulty is that we have no way to understand
the differing behaviors from a diagrammatic point of view.

Note that the situation is qualitatively different from our
scenario for isospin matter at $ \mu > m_\pi/2$.  In isospin
matter the phase transition at $\mu= m_\pi/2$ implies a new
$1/N_c$ expansion, and associated with this new expansion is a new
set of diagrams:  those with more than a single quark loop.
However, an analogous behavior cannot happen for quenched
QCD---by construction, in quenched QCD only single quark loop
graphs can contribute.   In this diagrammatic sense, large $N_c$
quenched QCD with a baryon chemical potential is {\it not}
analogous to  large $N_c$ QCD with an isospin chemical potential.

The upshot of this is that the difference between full and
quenched QCD at large $N_c$ and non-zero baryon chemical potential
is not fully understood.  The argument analogous to the one used
for the isospin case does not resolve the issue at a diagrammatic
level.  Until this issue is fully resolved one must view the
scenario presented in Sec.~\ref{PR} as being  provisional. While
the scenario seems quite plausible when viewed in its own right,
the lack of understanding the apparently analogous problem in QCD
suggests the possibility of a common explanation for both issues
which is presently unknown.  Of course, it is also possible that
the scenario in Sec.~\ref{PR} is, in fact, correct and that there
are subtle issues associated with quenching which will ultimately
explain the puzzle discussed in this section. This is, perhaps,
not too implausible. Quenched QCD, after all, is not a true
physical theory and it is quite conceivable that some peculiar
feature of it can explain the diagrammatic issue raised in this
section while leaving the scenario of Sec.~\ref{PR} intact.
Clearly, however, new ideas are needed to resolve this issue and
clarify the nature of large $N_c$  QCD with non-zero chemical
potentials.

 \acknowledgments  This work was supported by the
U.S.~Department of Energy through grant DE-FG02-93ER-40762.

\end{document}